\documentclass[twocolumn]{aastex631}
\usepackage{bm}
\usepackage{epsfig}
\usepackage[utf8]{inputenc}
\usepackage{ulem}
\usepackage{graphicx}
\usepackage{color}
\usepackage{subfigure}
\newcommand{\be}{\begin{equation}}
\newcommand{\ee}{\end{equation}}
\newcommand{\beq}{\begin{equation}}
\newcommand{\eeq}{\end{equation}}
\newcommand\bea{\begin{eqnarray}}
\newcommand\eea{\end{eqnarray}}

\shorttitle{spectra of magnetic turbulence}
\shortauthors{Vega et al.}

\begin{document}


\title{Spectra of magnetic turbulence in a relativistic plasma}

\correspondingauthor{Cristian Vega}
\email{csvega@wisc.edu}

\author{Cristian Vega}
\affiliation{Department of Physics, University of Wisconsin at Madison, Madison, Wisconsin 53706, USA}

\author[0000-0001-6252-5169]{Stanislav Boldyrev}
\affiliation{Department of Physics, University of Wisconsin at Madison, Madison, Wisconsin 53706, USA}
\affiliation{Center for Space Plasma Physics, Space Science Institute, Boulder, Colorado 80301, USA}

\author[0000-0003-1745-7587]{Vadim Roytershteyn}
\affiliation{Center for Space Plasma Physics, Space Science Institute, Boulder, Colorado 80301, USA}




\date{\today}

\begin{abstract}
We present a phenomenological and numerical study of strong Alfv\'enic turbulence in a magnetically dominated collisionless relativistic plasma with a strong background magnetic field.  In contrast with the non-relativistic case, the energy in such turbulence is contained in magnetic and electric fluctuations. We argue that such turbulence is analogous to turbulence in a strongly magnetized non-relativistic plasma in the regime of broken quasi-neutrality. 
Our 2D particle-in-cell numerical simulations of turbulence in a relativistic pair plasma find that the spectrum of the total energy has the scaling $k^{-3/2}$, while the difference between the magnetic and electric energies, the so-called residual energy, has the scaling $k^{-2.4}$. The electric and magnetic fluctuations at scale $\ell$ exhibit dynamic alignment with the alignment-angle scaling close to $\cos\phi_\ell\propto \ell^{1/4}$. At scales smaller than the (relativistic) plasma inertial scale, the energy spectrum of relativistic inertial Alfv\'en turbulence steepens to $k^{-3.5}$. 
\end{abstract}



\section{Introduction}

Relativistic plasma turbulence may be present in astrophysical objects, such as jets from active galactic nuclei, pulsar-wind nebulae, magnetospheres of stars and accretion discs. In particular, interest has been attracted to the magnetically-dominated regime, where the magnetic energy exceeds the rest-mass energy of the plasma. It has been discovered that such turbulence leads to efficient energization of plasma particles. This leads not only to thermal plasma heating but also to particles accelerating to ultrarelativistic energies, a process manifested in power-law tails of the particle energy distribution functions \cite[][]{zhdankin2017a,zhdankin2018,zhdankin2018c,zhdankin2019,zhdankin2021,comisso2018,comisso2019,comisso2021,zhdankin2020,nattila2020,demidem2020,vega2021,chernoglazov2021,nattila2021}. In this respect, magnetically dominated relativistic turbulence can be considered as a mechanism of particle acceleration complementary to previously studied particle acceleration by collisionless shocks or magnetic-reconnection  events \cite[e.g.,][]{marcowith2016,guo2020}.  

Numerical studies indicate that particle energization crucially depends on the properties of relativistic plasma turbulence itself. Recent numerical and phenomenological consideration suggests that 
 initially ultrarelativistic large-scale (outer-scale) turbulent fluctuations tend to become mildly relativistic in a few turnover times, whereas most of the energy of large-scale fluctuations is converted into thermal energy of plasma particles so that plasma temperature becomes ultrarelativistic. Moreover, the magnetic fluctuations exhibit Kolmogorov-like power-law spectra, bearing some similarity with the non-relativistic case \cite[e.g.,][]{vega2021,zhdankin2018c,comisso2019}. 
 
 Let us assume that a plasma is immersed in a uniform background magnetic field ${\bf B}_0$ and denote the typical (rms) strength of magnetic fluctuations as $\delta B_0$.  {The magnetization parameter related to the magnetic guide field, characterizes the ratio of the magnetic energy to the kinetic energy of plasma particles. In the relativistic case, it may be estimated as
$\sigma \sim {B_0^2}/{(4\pi n \gamma m c^2)}$,
where $n$ is the average number density of plasma particles, $m$ is their mass, and $\gamma$ is the typical gamma factor of their thermal distribution.} 
In this work, we concentrate on the regime $\sigma \gg 1$ and $B_0 \gg \delta B_0$. This may correspond to the case when a strong guide field is imposed by external sources (e.g, magnetoshperes), or when a small subregion of a turbulent domain is considered where magnetic fluctuations are much smaller than the mean magnetic field.  


Recent numerical and analytical studies of relativistic plasma turbulence using  magnetohydrodynamic (MHD) approximation suggested analogies with the non-relativistic case, such as the description in terms of the Elsasser fields, the similarities in the power-law exponents of the spectra of magnetic fluctuations, and the existence of  dynamic alignment between magnetic and velocity fields \cite[][]{tenbarge2021,chernoglazov2021}. It was also noted that the relative magnitude of the electric field fluctuations in this case exceeds that of the kinetic fluctuations \cite[][]{vega2021,chernoglazov2021}. 

{In this work, we study turbulence in a magnetically dominated relativistic collisionless plasma. We derive two-fluid equations governing dynamics of Alfv\'enic fluctuations and compare their predictions to first-principle kinetic particle-in-cell (PIC) simulations. We demonstrate that there is a remarkable mathematical correspondence of the derived equations with those describing dynamics of a {non-relativistic} plasma, which suggests that a similarity exists between properties of  turbulence in relativistic in non-relativistic plasmas. Such a similarity may look rather counter-intuitive from the physical point of view though, since in classical non-relativistic magnetohydrodynamic turbulence, electric fluctuations are negligible while magnetic fluctuations are approximately equipartitioned with fluid motions. In stark contrast with non-relativistic cases, the turbulence discussed here is energetically dominated by fluctuations of electromagnetic field alone.} 

{To resolve this seeming controversy, we note that non-relativistic plasma dynamics may, in fact, also contain a regime where electric field fluctuations are relatively strong as compared to kinetic fluctuations. Such a regime corresponds to broken plasma quaisneutrality, and it is rarely discussed in non-relativistic literature due to its limited applicability to practical systems. However, it turns out to be valuable for understanding magnetically dominated relativistic turbulence. The revealed physical similarity between relativistic and non-relativistic cases allows us to address the total energy spectrum of relativistic turbulence (the sum of electric and magnetic spectra) in both Alfv\'en and inertial Alfv\'en intervals, the scale-dependent angular alignment between electric and magnetic fluctuations, and the spectrum of the residual energy (the difference between the magnetic and electric spectra).}

{
\section{Alfv\'en dynamics of magnetically dominated plasma}
Consider non-relativistic motion of a relativistically hot collisionless plasma with temperarure $T_\alpha\gg m_\alpha c^2$, where $\alpha=\{e, i\}$ denotes the types of particles. The momentum equation takes the form \cite[e.g.,][]{mihalas1984}:
\begin{eqnarray}
\frac{\partial {\bm v}_\alpha}{\partial t}+\left({\bm v}_\alpha\cdot \bm{\nabla} \right){\bm v}_\alpha= \frac{-\bm{\nabla}p_\alpha}{\epsilon_\alpha/c^2}+
\frac{n_\alpha q_\alpha}{\epsilon_\alpha /c^2}\left[\bm{E}+\frac{\bm{v}_\alpha}{c}\times \bm{B} \right],\quad
\end{eqnarray}
{where $\epsilon_\alpha=n_\alpha m_\alpha c^2 + u_\alpha + p_\alpha$ is the enthalpy density, $n_\alpha$ is the particle number density, $u_\alpha$ is the internal energy density, and $p_\alpha$ is the pressure of the particles. For simplicity, we assume isotropic pressure. For untrarelativistic temperatures of the species, we can approximate $\epsilon_\alpha\approx u_\alpha +p_\alpha\approx 4p_\alpha$. In addition, one can assume some equation of state, for instance, the adiabatic law of relativistically hot plasma,  $p_\alpha\propto n_\alpha^{4/3}$.} 
}

{
We now assume that the fluctuations of magnetic and electric fields are much weaker than the large-scale uniform magnetic field ${\bm B}_0=B_0{\hat z}$.  We will be interested in Alfv\'enic plasma fluctuations that are relatively low frequency as compared to cyclotron frequencies, $\omega \ll \Omega_\alpha/\gamma_\alpha$, where $\gamma_\alpha$ is the typical gamma factor of particle thermal motion and $\Omega_\alpha$ is the non-relativistic gyrofrequency. We also impose a self-consistent assumption that the Fourier spectra of the fields are anisotropic in the Fourier space with respect to the background magnetic field ${\bm B_0}$, $k_\|\ll k_\perp$, and the fluctuations obey the critical balance condition $\delta B/B_0  \sim k_\|/k_\perp\ll 1$ \cite[e.g.,][]{goldreich_toward_1995}.
}

{
The following derivation is analogous to the procedure developed for the non-relativistic case \cite[e.g.,][]{chen_boldyrev2017,loureiro2018,boldyrev2021,milanese2020}. Here, we outline its main steps, the details can be found in the above references. As follows from the momentum equation, to the leading order in the small parameter $\omega\gamma_\alpha/\Omega_\alpha$ the particle motion is the E-cross-B drift, while to the next order it is the polarization drift, 
\begin{equation}
{\bm v}_{\alpha, \perp} = {\bm v}_E +\frac{\epsilon_\alpha}{n_\alpha q_\alpha B^2 c}\,{\bm B}\times \frac{d_E}{dt}{\bm v}_E,
\label{vperp}
\label{eq:drift}
\end{equation}
where ${\bm v}_{E}=c({\bm E}\times{\bm B})/B^2$ and the time derivative is $d_E/dt\equiv \partial/\partial t+{\bm v}_E\cdot{\bm \nabla}$.\footnote{We did not include the diamagnetic drift proportional to $\hat z\times \bm{\nabla} p_\alpha$ in the perpendicular velocity (\ref{eq:drift}), since such a drift does not lead to particle transport and should not contribute to the continuity equation.} The field-parallel component of the velocity field is expressed through the parallel electric current,
$n_\alpha v_{\alpha, \|}={J_{\alpha, \|}}/{q_\alpha}$.
These velocities are substituted into the continuity equation, 
\begin{eqnarray}
\frac{\partial n_\alpha}{\partial t}+\bm{\nabla}_\perp\cdot \left(n_\alpha \bm{v}_{\alpha, \perp} \right)+\nabla_\|(n_\alpha v_\|)=0.
\end{eqnarray}
To the leading order in magnetic, electric, and density fluctuations, one then obtains:
\begin{eqnarray}
\frac{\partial}{\partial t}\left(\frac{\delta n_\alpha}{n_0}-\frac{\delta B_z}{B_0}-\frac{\epsilon_{\alpha, 0}}{n_0 q_\alpha B_0^2}\nabla_\perp^2\phi\right)\nonumber \\
+\frac{1}{B_0}\left({\hat z}\times \bm{\nabla} \phi \right)\cdot \bm{\nabla} \left(\frac{\delta n_\alpha}{n_0}-\frac{\delta B_z}{B_0}-\frac{\epsilon_{\alpha, 0}}{n_0  q_\alpha B_0^2}\nabla_\perp^2\phi\right)\nonumber \\
+\frac{1}{n_0 q_\alpha}\nabla_\|J_{\alpha, \|}=0.\quad
\label{eq:A}
\end{eqnarray}
In this equation, $n_0$ is the unperturbed density of each species, $\delta n_\alpha=n_\alpha -n_0$ is the corresponding density perturbation, and $\phi$ is the electric potential, ${\bm E}=-\bm{\nabla}\phi-\frac{1}{c}\partial\bm{A}/\partial t$. The parallel gradient is taken along the direction of the magnetic field, $
\nabla_\|=\partial/\partial z-\frac{1}{B_0}({\hat z}\times \bm{\nabla} A_z)\cdot \bm{\nabla} $, where the field-perpendicular magnetic perturbation is expressed as $\delta {\bm B}_\perp=-{\hat z}\times \bm{\nabla} A_z$.  
In order to find the electric current, we turn to the magnetic-field-parallel component of the momentum equation:
\begin{eqnarray}
\frac{\partial {v}_{\alpha, \|}}{\partial t}+\left({\bm v}_E\cdot \bm{\nabla}_\perp \right){\bm v}_{\alpha, \|}= \frac{-{\nabla}_\| p_{\alpha}}{\epsilon_\alpha/c^2}+
\frac{n_\alpha q_\alpha}{\epsilon_\alpha /c^2}{E}_\|.\quad
\label{eq:B}
\end{eqnarray}
}
{
We now multiply each of the Equations (\ref{eq:A}) and (\ref{eq:B}) by $n_0 q_\alpha$ and sum over particle species. As a result, Equation (\ref{eq:A}) turns into the charge conservation law:
\begin{eqnarray}
\frac{\partial}{\partial t}\left({\rho}-\frac{2\epsilon_{0}}{ B_0^2}\nabla_\perp^2\phi\right)
+\frac{1}{B_0}\left({\hat z}\times \bm{\nabla} \phi \right)\cdot \bm{\nabla} \left({\rho}-\frac{2\epsilon_{0}}{  B_0^2}\nabla_\perp^2\phi\right)\nonumber \\
+\nabla_\|J_{\|}=0,\quad\quad
\label{eq:A2}
\end{eqnarray}
where $\rho=q_i\delta n_i+q_e\delta n_e$ is the electric charge density, $J_\|=J_{e, \|}+J_{i, \|}$ is the parallel current, and $\epsilon_0=\left(\epsilon_{i, 0}+\epsilon_{e, 0}\right)/2$ is the mean unperturbed enthalpy.  
To simplify the formulae, we have assumed without loss of generality that $q_i=|q_e|\equiv q$. We may also assume that in the relativistic case, the unperturbed enthalpy is the same for both species, $\epsilon_{i,0}=\epsilon_{e,0}= \epsilon_0$. In the ultrarelativistic limit, $\epsilon_0=4p_0$. The parallel momentum equation (\ref{eq:B}) will then lead to
\begin{eqnarray}
\frac{\partial {J}_\|}{\partial t}+\left({\bm v}_E\cdot \bm{\nabla}_\perp \right){J}_{\|}= \frac{- q n_0c^2}{\epsilon_0}{\nabla}_\| \delta p +
\frac{2 n_0^2 c^2 q^2}{\epsilon_0}E_\|,\quad
\label{eq:B2}
\end{eqnarray}
where $\delta p=\delta p_i-\delta p_e$ denotes the pressure imbalance. 
}

{
We note \cite[see also][]{vega2021} that the terms containing the electric charge density $\rho$ and pressure imbalance $\delta p$ in Eqs.~(\ref{eq:A2}, \ref{eq:B2}) reflect deviation of plasma fluctuations from quasineutrality, that is, from the condition $\delta n_i=\delta n_e$. It is easy to see by using the Gauss law, $-\nabla^2\phi=4\pi \rho$, that these terms are relatively small in the case of weak magnetization, that is, when $\sigma \sim B_0^2/(8\pi \epsilon_0)\ll 1$. In the opposite case of strong magnetization that we consider in this work, the deviation from quasineutrality is essential and as a result, both the electric charge and pressure imbalance become dynamically significant.  In the case of a non-relativistic electron - proton plasma, we need to replace  $\epsilon_0\to n_0m_ic^2/2$, and the magnetization parameter turns into the so-called quasineutrality parameter, $\Omega_i^2/\omega^2_{pi}=\lambda_i^2/\rho_i^2$. Here, $\lambda_i$ is the ion Debye scale, $\rho_i$ is the gyroscale, and 
$\omega_{pi}$ is the ion plasma frequency.   Therefore, $\Omega_i^2/\omega^2_{pi}$ is the non-relativistic analog of relativistic magnetization~$\sigma$. 
}

{
In order to close system (\ref{eq:A2})-(\ref{eq:B2}), we replace the charge density by using the Gauss law, $\rho=-(1/4\pi)\nabla_\perp^2\phi$, express the parallel electric current as $J_\|=-(c/4\pi)\nabla_\perp^2 A_z$, and use the adiabatic law for each particle species to evaluate the pressure gradients, so that
\begin{eqnarray}
\nabla_\|\delta p=\frac{4}{3}\frac{p_0}{n_0}\nabla_\|\left(\delta n_i-\delta n_e\right) =-\frac{\epsilon_0}{3 n_0}\frac{1}{4\pi q}\nabla_\| \nabla^2_\perp \phi.\quad
\end{eqnarray}
Here, we assume that the unperturbed pressure is the same for both species, $p_0=p_{i,0}=p_{e,0}$. Finally, we introduce the new variables for the scalar and vector potentials, ${\tilde \phi}={\phi} c/B_0$, ${\tilde A}_z={A}_z c/\sqrt{8\pi \epsilon_0}$. Below, we will use only these variables and omit the overtilde sign. Substituting these expressions into Eqs.~(\ref{eq:A2}) and (\ref{eq:B2}), we finally obtain the system of equations describing Alfv\'en dynamics of a relativistic plasma in both magnetohydrodynamic and inertial regimes:
\begin{eqnarray}
\frac{\partial}{\partial t}\nabla^2_\perp\phi 
+\left(\hat{{z}}\times \bm{\nabla}_{\perp} \phi \right)\cdot \bm{\nabla}_{\perp} \nabla^2_\perp\phi \quad  \nonumber \\
=-\frac{v_A}{1+v_A^2/c^2}\nabla_{\|} \nabla_{\perp}^{2} A_z, \quad \label{low_beta}\\
 \frac{\partial}{\partial t} \left( A_z -d^2\nabla^2_\perp A_z\right) - \left(\hat{{z}}\times \bm{\nabla}_\perp\phi\right)\cdot\bm{\nabla}_\perp d^2\nabla^2_\perp {A_z} \quad \nonumber \\
 = - v_A \nabla_\| \left( {\phi} -\frac{1}{3}d^2\nabla_\perp^2\phi \right),\quad
\label{reduced} 
\end{eqnarray}
where $v_A=c B_0/\sqrt{(8\pi \epsilon_0)}$ is the Alfv\'en speed,\footnote{{In relativistic studies, it is common to use the relativistic Alfv\'en speed defined as ${\tilde v_A}=v_A/\sqrt{1+v_A^2/c^2}$, so that it is always bounded by the speed of light. In our discussion, we use the quantity $v_A$ to make the analogy with the non-relativistic case more transparent. The magnetically dominated case considered in this work corresponds to ${\tilde v_A}\approx c$.}}. $d=\sqrt{\epsilon_0/(8\pi n_0^2q^2)}$ is the relativistic inertial length, and the field-parallel gradient has the form 
\begin{eqnarray}
\nabla_\|=\partial/\partial z-\frac{1}{v_A}({\hat z}\times \bm{\nabla}_\perp A_z)\cdot \bm{\nabla}_\perp .
\end{eqnarray}
}
{
The dispersion relation of the linear waves described by Eqs.~(\ref{low_beta}, \ref{reduced}) is
\begin{eqnarray}
\omega^2=\frac{k_z^2 v_A^2}{1+v_A^2/c^2}\frac{1+d^2k_\perp^2/3}{1+d^2k_\perp^2},
\label{iaw}
\end{eqnarray}
which can be termed the {\it relativistic inertial Alfv\'en} waves. 
}

{
Except for the very last term in Eq.~(\ref{reduced}) describing the relativistic pressure contribution, the system of equations (\ref{low_beta}) and 
(\ref{reduced}) is analogous to the non-relativistic case. The nonrelativistic electron-ion case is recovered by replacing the Alfv\'en speed and the inertial length by their non-relativistic counterparts by means of the substitutions $\epsilon_0\to n_0m_ic^2/2$ in the Alfv\'en velocity and $\epsilon_0\to 2n_0m_ec^2$ in the inertial length. It may also be instructive to compare the dispersion relation (\ref{iaw}) with the dispersion relation of non-relativistic inertial kinetic Alfv\'en waves \cite[e.g.,][]{streltsov95,lysak96}, \cite[][Eq. 19]{boldyrev2021} where, similarly, the kinetic correction coming from thermal particle motion enters the numerator, while the inertial correction appears in the denominator. In our relativisic case, these two corrections are necessarily on the same order.  
We also note that similarly to the previous study \cite[][]{tenbarge2021}, at large scales $k_\perp^2d^2\ll 1$, Eqs. (\ref{low_beta}, \ref{reduced}) describe shear Alfv\'en waves in a relativistically hot plasma and they are mathematically analogous to the equations of reduced magnetohydrodynamics.
}

{
Finally, as can be checked directly, Eqs.~(\ref{low_beta}, \ref{reduced}) conserve two quadratic integrals, the energy:
\begin{eqnarray}
W=\frac{\epsilon_0}{c^2}\int \bigg\{\left(\nabla_\perp A_z\right)^2+d^2\left(\nabla_\perp^2 A_z\right)^2  \nonumber \\
+\left[\left(\nabla_\perp \phi\right)^2+\frac{d^2}{3}\left(\nabla_\perp^2\phi\right)^2\right]\left(\frac{v_A^2}{c^2}+1\right) \bigg\}\,d^3 x
\label{energy}
\end{eqnarray}
and the generalized cross-helicity:
\begin{eqnarray}
H=\frac{\sqrt{\epsilon_0}}{c}\int {\nabla}^2_\perp \phi \left(A_z-{d^2}\nabla_\perp^2 A_z \right)\,d^3 x.
\end{eqnarray}
}

\section{Numerical results}
In the hydrodynamic range of scales, $k_\perp^2 d^2\ll 1$,  the energy integral becomes:
\begin{eqnarray}
W=\frac{\epsilon_0}{c^2}\int\left[(\nabla_\perp A_z)^2+(\nabla_\perp \phi)^2\left(\frac{v_A^2}{c^2}+1\right) \right]d^3 x \nonumber \\
=\int \left[\frac{(\delta B)^2}{8\pi}+ \frac{E^2}{8\pi} +\epsilon_0\frac{v_E^2}{c^2}\right]d^3 x,\quad
\end{eqnarray}
where in the second expression, we have restored the corresponding physical fields. 
We see that the ratio of the electric to kinetic energy is given by the parameter $v_A^2/c^2=B_0^2/(8\pi \epsilon_0)\sim \sigma $. In both relativistic and non-relativistic cases, when $\sigma \gg 1$, the charge fluctuations are significant and the electric energy dominates the kinetic energy. 
Therefore, the energy of fluctuations is mostly contained in magnetic and electric fields. 

{In this limit, the system of equations (\ref{low_beta}, \ref{reduced}) is mathematically analogous to the equations of non-relativistic reduced magnetohydrodynamics, with the only difference that in the magnetically dominated case, the field $\phi$ in these equations should be associated with the intensity of electric rather than kinetic fluctuations.} Based on this analogy, we may conjecture that the spectrum of the total energy of relativistic magnetically dominated plasma turbulence, the spectrum of its residual energy, and the alignment angle of turbulent  fluctuations, should be similar to their reduced MHD counterparts when reinterpreted in terms of magnetic and electric fields as discussed above. Here, we analyze these spectra using kinetic particle-in-cell simulations of a relativistic collisionless plasma.   

We numerically study decaying 2D turbulence in a pair plasma where the imposed uniform magnetic field ${\bf B_0}=B_0{\hat z}$ is much stronger than the initial magnetic perturbations. We use the fully relativistic particle-in-cell  code VPIC \cite[][]{bowers2008}. The fluctuating fields (magnetic, electric, particle velocities) have three vector components, but depend on only two spatial coordinates $x$ and $y$.  Similarly to our previous study \cite[][]{vega2021}, we denote the root-mean-square of initial magnetic perturbations as $\delta B_{0}=\langle\delta B^2({\bf x}, t=0)\rangle^{1/2}$,
and define the two magnetization parameters related to the guide magnetic field and the fluctuations, respectively: 
\begin{eqnarray}
\sigma_0=\frac{B_0^2}{4\pi n_0w_0mc^2}, \quad
{\tilde \sigma}_0=\frac{(\delta B_0)^2}{4\pi n_0w_0mc^2}.
\end{eqnarray}
In these formulae,  $n_0$ denotes the mean density of {\em each} plasma species, $w_0 mc^2$ is the initial particle enthalpy, with 
$w_0=K_3(1/\theta_0)/K_2(1/\theta_0)$. 
Here $K_\nu $ is the modified Bessel function of the second kind, and we use the dimensionless initial temperature parameter $\theta_0=kT_0/mc^2$. The particle distribution function is initialized with an isotropic Maxwell-J\"uttner distribution, with the temperature parameter $\theta_0=0.3$. For such an initialization, $w_0\approx 1.88$.

We use a double-periodic square simulation domain with dimensions $L_x=L_y \approx 2010 \,d_e$, where $d_e$ is the (nonrelativistic) inertial scale of the plasma particles. The adopted numerical resolution is $N_x=N_y=16640$. The simulations have 100 particles per cell per species. The simulation plane is normal to the uniform magnetic field. The runs are initiated by randomly phased magnetic fluctuations of the Alfv\'enic type
\begin{eqnarray}
\delta\mathbf{B}(\mathbf{x})=\sum_{\mathbf{k}}\delta B_\mathbf{k}\hat{\xi}_\mathbf{k}\cos(\mathbf{k}\cdot\mathbf{x}+\chi_\mathbf{k}),
\end{eqnarray}
where the wave numbers are chosen in the interval $\mathbf{k}=\{2\pi n_x/L_y,2\pi n_y/L_y\}$, with $n_x,n_y=1,...,8$, and $\chi_k$ are the random phases. The field polarizations correspond to the shear-Alfv\'en modes \cite[e.g.,][]{lemoine2016,demidem2020}, $\hat{\xi}_\mathbf{k}=\mathbf{k}\times\mathbf{B}_0/|\mathbf{k}\times\mathbf{B}_0|$, and all the amplitudes $\delta B_{\mathbf{k}}$ are chosen to be the same.

\begin{table}[tb!]
\centering
\vskip0mm
\hskip0.0cm \begin{tabular}{c c c c c c c} 
\hline
Run & $\sigma_0$ & ${{\tilde \sigma}_0 }$ & $\left({B_0}/{\delta B_0}\right)^2$ &  $\omega_{pe}\delta t$ & $\langle\gamma\rangle$  & $\langle\tilde\gamma\rangle$    \\
\hline
I & 90 & 10 & 9 & 0.02 & 3.30 & 1.04\\
II & 360 & 40 & 9 & 0.02 & 8.37 & 1.08\\
\hline
\end{tabular}
\caption{Parameters of the runs (magnetization, magnetic field strengths, time steps), as well as the averaged particle Lorentz factors $\gamma$ and fluid Lorentz factors $\tilde{\gamma}$. The time steps are normalized to the nonrelativistic single-species plasma frequency $\omega_{pe}$.}
\label{table}
\end{table}
Table~{\ref{table}} summarizes the simulations discussed in this paper. As was previously noted in \cite[][]{vega2021}, the bulk turbulent fluctuations are only mildly relativistic, with average Lorentz factor $\langle\tilde\gamma\rangle\gtrsim 1$, while the particles are strongly relativistic, with average Lorentz factor $\langle\gamma\rangle\approx3.3-8.4$. This is consistent with the phenomenological discussion of \cite[][]{vega2021} that  as the initially strong magnetic fluctuations relax, the magnetic energy is mostly converted into heat rather than kinetic energy of collective plasma motion. The Lorentz factors were measured at simulation time $t=16l/c$, or about two light crossing times of the simulation box, where $l=2\pi/k_{x,y}\,(n=8)=L_{x,y}/8$.
By this time, quasi-steady states for the distributions of fields have been established. The light crossing time of the simulation box corresponds to a few large-scale dynamical times of turbulence, $l/v$, where $v\approx 0.3 - 0.4\, c$ is the typical velocity of turbulent fluctuations.

\begin{figure}[tb!]
\includegraphics[width=\columnwidth,height=0.55\columnwidth]{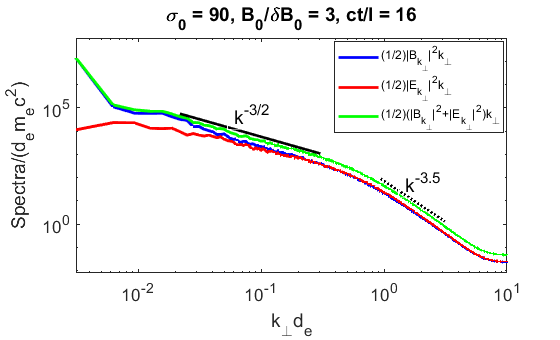}
\includegraphics[width=\columnwidth,height=0.55\columnwidth]{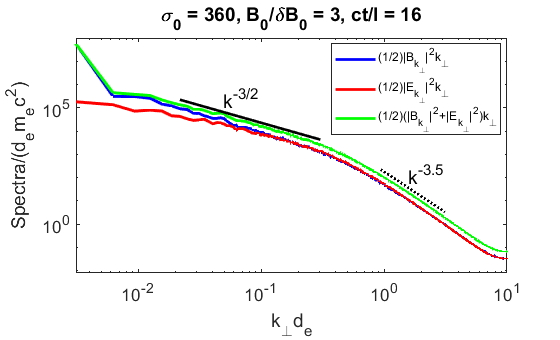}
\caption{The magnetic and electric spectra, and the total electromagnetic energy spectrum, for two different magnetizations. The total energy spectrum approaches a $k^{-3/2}$ power-law at $k_\perp d_e\ll 1$, and a $k^{-3.5}$ power-law at $k_\perp d_e\gg 1$.}
\label{spectra}
\end{figure}

In Figure \ref{spectra}, we present the spectra of magnetic and electric fluctuations as well as the total energy spectrum, $W_{k_\perp}=|B_{k_\perp}|^2+|E_{k_\perp}|^2$. The phase-volume compensated scaling of the energy spectrum in the magnetohydrodynamic interval of scales $k_\perp d_e\ll 1$ is close to $W_{k_\perp}2\pi k_\perp \propto k_{\perp}^{-3/2}$. Such a spectrum is expected in non-relativistic magnetohydrodynamic turbulence \cite[e.g.,][]{boldyrev2006,boldyrev2009,mason2006,mason2012,perez_etal2012,tobias2013,chandran_intermittency_2015,chen2016,kasper2021}, where the energy is contained in magnetic and kinetic fluid fluctuations. We see that it also holds in relativistic collisionless plasma turbulence dominated by magnetic and electric fields. Our result is also consistent with the recent relativistic MHD studies \cite[][]{tenbarge2021,chernoglazov2021}.

{
At kinetic scales, $1/d_e^2\ll k_\perp^2 \ll 1/\rho_e^2$, the inertial Alfv\'en waves are transformed into $\omega^2=\frac{1}{3}k_z^2v_A^2/(1+v_A^2/c^2)$, while the energy integral (\ref{energy}) takes the form: 
\begin{eqnarray}
W=\frac{\epsilon_0}{c^2}\int\left[ \left(\nabla_\perp^2 A_z\right)^2 
 +\frac{1}{3}\left(\nabla_\perp^2\phi\right)^2\left(\frac{v_A^2}{c^2}+1\right) \right]d^3 x. \qquad
\label{energy_kin}
\end{eqnarray}
In this asymptotic region, dimensional estimates give for turbulent fluctuations at field-perpendicular scale $\ell$: $A_{z,\ell}\sim \phi_\ell$, and the nonlinear interaction time is estimated as $\tau_\ell\sim \ell^2/\phi_\ell$. The condition of constant flux of the conserved quantity (\ref{energy_kin}) then reads $\phi_\ell^2/(\ell^4 \tau_\ell)=\rm{const}$, which gives for the scaling of fluctuations $\phi_\ell\propto \ell^2$ and for the electromagnetic energy spectrum of relativistic inertial Alfv\'en waves $W_{k_\perp}2\pi k_\perp\propto k_\perp^{-3}$. The nonlinear interaction time for such modes turns out to be independent of scale, which implies that for the critically balanced cascade, $\omega \propto 1/\tau_\ell$, we have $k_z\propto \rm{const}$, that is, the cascade proceeds in the field-perpendicular direction. 
}

{
We note that such a turbulent cascade is somewhat analogous to the cascade of enstrophy in incompressible non-relativistic two-dimensional fluid turbulence. Such a cascade is however only marginally local, so it depends on the conditions at the low-$k$ boundary of the inertial interval. Numerical simulations and experiments typically reveal the spectra steeper than the $-3$ in this case, unless the formation of large-scale structures is controlled (suppressed) by some forcing mechanism and/or the inertial interval is sufficiently large \cite[e.g.,][]{boffetta2012}. Our numerical simulations found the spectral scaling close to $k_\perp^{-3.5}$. In addition to the limited separation between the particle inertial and gyroradius scales and possible spatial intermittency, in our case the steepening may also be related to Landau damping at kinetic scales \cite[e.g.,][]{nattila2021} as the phase velocity of the inertial-Alfv\'en waves, $c/\sqrt{3}$, is smaller than the thermal speed of particles, $\sim c$.}

\begin{figure}[tb!]
\includegraphics[width=\columnwidth,height=0.55\columnwidth]{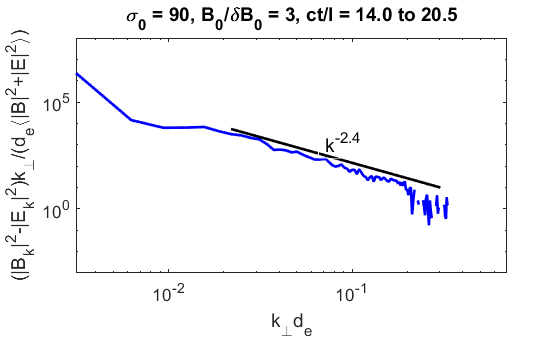}
\includegraphics[width=\columnwidth,height=0.55\columnwidth]{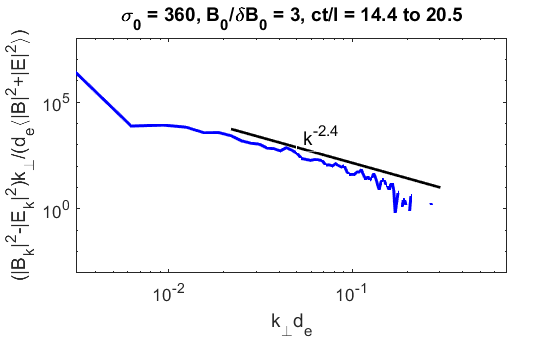}
\caption{The spectra of the normalized residual energy for two different magnetizations, averaged over nine data cubes covering the indicated  intervals of turn-over times. The solid lines are shown for reader's reference. }
\label{residual}
\end{figure}

In non-relativistic magnetohydrodynamic turbulence, the magnetic energy is known to exceed the energy of kinetic fluctuations. Phenomenological and numerical considerations demonstrated that the difference between the magnetic and kinetic energies, the so-called residual energy, is positive and has the spectrum close to $-2$ \cite[e.g.,][]{boldyrev2011,boldyrev2012,chen2013}. In the relativistic magnetically dominated case, we may introduce the analog of residual energy as the difference between the magnetic and electric energies, $R_{k_\perp}=|B_{k_\perp}|^2-|E_{k_\perp}|^2$. This quantity is measured in  Figure~\ref{residual}. In order to compensate for overall energy decline in decaying turbulence we have normalized the residual energy by the total energy of fluctuations and then averaged over several data cubes. A power-law spectrum is indeed observed, however, the scaling is slightly steeper than in its non-relativistic counterpart,  more consistent with $R_{k_\perp}2\pi k_\perp\propto k_{\perp}^{-2.4}$. 

\begin{figure}[tb!]
\includegraphics[width=\columnwidth,height=0.55\columnwidth]{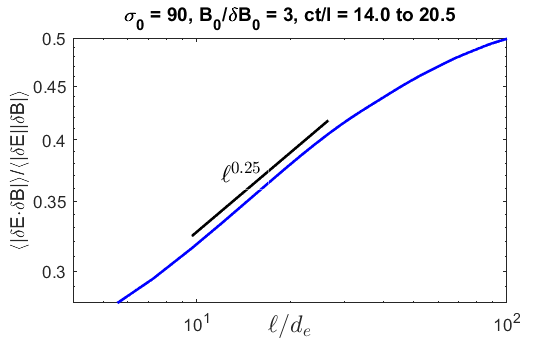}
\includegraphics[width=\columnwidth,height=0.55\columnwidth]{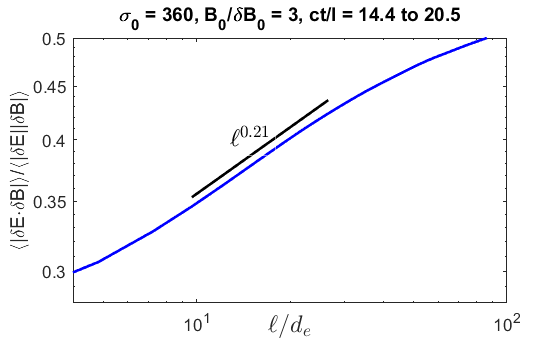}
\caption{The cosine of the alignment angle $\phi_\ell$ between the magnetic and electric fluctuations in numerical simulations. The results are averaged over nine data cubes covering the indicated time intervals. The straight lines are shown for reference.}
\label{alignment}
\end{figure}
Finally, a characteristic feature of non-relativistic magnetic plasma turbulence in the presence of a strong guide field, is the dynamic alignment between the shear-Alfv\'en magnetic and velocity fluctuations. Such fluctuations become spontaneously aligned in a turbulent cascade, with the alignment angle scaling as $\sin\theta_\ell\propto\ell^{1/4}$, \cite[e.g.,][]{boldyrev2006,podesta_scale_2009,chen_11,perez_etal2012}. Such an alignment progressively reduces the strength of nonlinear interaction at small scales, which arguably explains the shallower-than-Kolmogorov spectrum of turbulent energy,~$k_\perp^{-3/2}$.   In the case of magnetically-dominated turbulence, one may similarly expect a scale-dependent dynamic alignment between the electric and magnetic field fluctuations,
\begin{equation}\label{angle}
\cos\phi_\ell=\frac{\langle|\delta\mathbf{E}\cdot\delta\mathbf{B}|\rangle}{\langle|\delta\mathbf{E}||\delta\mathbf{B}|\rangle},
\end{equation}
where ${\ell}$ is the scale of the fluctuations in the field-perpendicular plane, e.g., $\delta {\bf B}\equiv {\bf B}({\bf x}+{\bf \bm{\ell}})-{\bf B}({\bf x})$ where $\bm{\ell}\perp {\bf B}_0$. In order to see whether a similar alignment exists in our numerical simulations of collisionless relativistic turbulence, 
we plot the cosine of the angle $\phi_\ell$ vs scale $\ell$ in Figure~\ref{alignment}. We observe a scaling close to that of the non-relativistic case, and also to the relativistic MHD case \cite[][]{chernoglazov2021}. 
We however notice that  the scaling varies slightly with the plasma magnetization parameter $\sigma_0$; it is slightly shallower in the case of stronger magnetization, possibly reflecting a shorter inertial range due to larger relativistic inertial scale.

\section{Discussion}

We have presented a numerical and phenomenological study of relativistic turbulence in a collisionless magnetically dominated plasma. We proposed that dynamic equations (\ref{low_beta}, \ref{reduced})  provide a useful phenomenological model for magnetically dominated relativistic plasma turbulence. Using the 2D particle-in-cell simulations, we demonstrated that in such turbulence, the energy is contained mostly in collective magnetic and electric fluctuations. In the magnehydrodynamic range of scales, $k_\perp^2 d_e^2\ll 1$, the total energy (magnetic plus electric) exhibits the spectrum close to that of Alfv\'enic turbulence, $W_{k_\perp}2\pi k_\perp\propto k_\perp^{-3/2}$. In the kinetic range, $1/d_e^2\ll k_\perp^2\ll 1/\rho_e^2$, the turbulence is governed by relativistic inertial Alfv\'en modes, and the spectrum steepens to approximately $k_\perp^{-3.5}$.  

We further established that in relativistic magnetically dominated turbulence, there is excess of magnetic energy over electric one, which becomes progressively smaller at smaller scales. We propose that this phenomenon is analogous to the generation of the so-called residual energy, that is, the excess of magnetic over kinetic energy known in the non-relativistic quasineutral case. The measured spectrum of the residual energy is close to  $R_{k_\perp}2\pi k_\perp \propto k_\perp^{-2.4}$, which is slightly steeper than its non-relativistic counterpart, indicating an interesting difference with the non-relativistic case.

An additional intriguing similarity of relativistic turbulence with the nonrelativistic case is manifested in the presence of the so-called  scale-dependent dynamic alignment between magnetic and electric fluctuations. We have found that such fluctuations become progressively more orthogonal to each other at smaller scales. The cosine of the angle between the fluctuations was found to scale close to $\cos\phi_\ell\propto \ell^{0.25}$. Based on the analogy with the nonrelativistic case, we proposed that such an alignment reduces the strength of nonlinear interactions in the relativistic dynamics, thus explaining the observed $-3/2$ scaling of the energy spectrum. We also note that the run with a stronger magnetization resulted in a slightly shallower scaling of the alignment angle. This may be related to a shorter inertial interval of turbulence due to a larger relativistic inertial scale of thermal particles.   

Finally, we note that our consideration may provide a useful framework for phenomenological studies of relativistic magnetically dominated plasma turbulence in a presence of a strong background magnetic field. In particular, it may be relevant for the analysis of particle heating and acceleration mediated by such turbulence, as well as the effects of magnetic reconnection, struture formation, and intermittency generated by a turbulent cascade.

\smallskip
The work of CV and SB was partly supported by NSF Grant PHY-2010098, by NASA Grant 80NSSC18K0646, and by the Wisconsin Plasma Physics Laboratory (US Department of Energy Grant DE-SC0018266). VR was partially supported by NSF/DOE Partnership in Basic Plasma Science and Engineering through grant DE-SC0019315 and by NASA grant 80NSSC21K1692. Computational resources were provided by the Texas Advanced Computing  Center at the University of Texas at Austin (XSEDE Allocations No. TG-PHY110016 and TG-ATM180015) and by the NASA High-End Computing (HEC) Program through the NASA Advanced Supercomputing (NAS) Division at Ames Research Center.


\bibliography{references}
\end{document}